\documentclass[fleqn,twoside]{article}
\usepackage{espcrc2}
\usepackage{graphicx}

\def\MG#1{[[MG]]}
\def\YS#1{[[YS]]}

\def\bar#1{\overline{#1}}
\def\tg{\tilde{g}}
\def\brs{\hat\d}
\def\half{{1\over 2}}
\def\eg{\mbox{\it e.g.} }

\def\beq{\begin{equation}}
\def\eeq{\end{equation}}
\def\bqry{\begin{eqnarray}}
\def\eqry{\end{eqnarray}}

\def\seeq#1{eq.~(\ref{#1})}

\def\seneq#1{~(\ref{#1})}

\def\rcite#1{ref.~\cite{#1}}

\def\CMP#1{Comm. Math. Phys. {\bf #1}}

\def\NPB#1{Nucl. Phys. {\bf B#1}}

\def\PLB#1{Phys. Lett. {\bf B#1}}
\def\PRD#1{Phys. Rev. {\bf D#1}}

\def\b{\beta}
\def\c{\chi}
\def\d{\delta}
\def\f{\phi}                    

\def\h{\eta}

\def\l{\lambda}
\def\m{\mu}

\def\x{\xi}

\def\L{\Lambda}
\def\O{\Omega}

\def\cd{{\cal D}}

\title{Non-perturbative BRST invariance and what it might be good for}

\author{Yigal Shamir\address{
School of Physics and Astronomy,
Tel-Aviv University, Ramat~Aviv 69978, ISRAEL}
and
Maarten Golterman\address{
Department of Physics and Astronomy,
San Francisco State University,
San Francisco, CA 94132, USA}}

\begin{document}

\begin{abstract}
We construct a local, gauge-fixed, lattice Yang-Mills theory
with an exact BRST invariance, and with the same perturbative
expansion as the standard Yang-Mills theory.
The ghost sector, and some of its BRST transformation rules,
are modified to get around Neuberger's theorem.
A special term is introduced in the action to regularize
the Gribov horizons, and the limit where the regulator is removed is discussed.
We conclude with a few comments on what might be the physical
significance of this theory. We speculate that there may exist
new strong-interaction phases apart from the anticipated confinement phase.
(For additional technical details see \rcite{http}.)

\end{abstract}

\maketitle

\section{Introduction}

The conventional lattice formulation of gauge theories
does not require gauge fixing because the link variables
take values in the compact gauge group.
At the same time, we have gained a lot of knowledge about gauge theories
from perturbation theory, where gauge fixing is indispensable.
This includes for example the celebrated asymptotic freedom
of non-abelian gauge theories. Given these facts, it is not unlikely
that we could learn new things if we had at our disposal
a local, non-perturbative ({\it i.e.} lattice)
formulation that makes closer contact
with standard perturbation theory, {\it including} its gauge-fixing sector.

Most common in perturbation theory
are covariant gauges for which power-counting
renormalizability is manifest.
Here we will focus on pure Yang-Mills (YM) theory
with the covariant gauge-fixing term $(\partial_\m A_\m)^2/(2\x g^2)$,
where $g$ is the gauge coupling and $\xi$ is the gauge parameter.
The gauge-fixed action contains also the ghost term
$\bar{C}_a \O_{ab} C_b$ where the
(real, but in general non-hermitian) Faddeev-Popov operator is
$\O_{ab} = \partial_\m D_{ab\m}$, and
$D_{ab\m}= \d_{ab}\partial_\m + f_{acb} A_{c\m}$.
The gauge-fixed action is invariant under  the BRST transformation.
The latter resembles a gauge transformation with a grassmannian parameter,
the ghost field $C(x)$, and plays a key role in proving
renormalizability and unitarity to all orders.

We may now state our goal by asking the following question:
{\it can one construct a local, gauge-fixed, BRST-invariant
lattice formulation of Yang-Mills theory?}
Success is by no means guaranteed, but
there are good reasons to try. Among our
motivations we list the following.

\noindent
1) A non-perturbative construction of {\it non-abelian chiral gauge theories}
is another long-standing open problem.
In its absence, our knowledge of the dynamics of chiral gauge theories
lags way behind our understanding of QCD. It has been demonstrated
convincingly that fermions with chiral coupling to an (abelian) gauge field
exist on the lattice if the action contains a covariant gauge-fixing
term~\cite{chgt}. Gauge fixing may thus prove to be a key element in the
lattice construction of non-abelian chiral gauge theories as well.

\noindent
2) A BRST-invariant gauge-fixed lattice theory might, possibly,
contain new phases except the familiar confinement phase
(see section 5).

\section{The Gribov problem}
By definition, a gauge-fixing term destroys gauge invariance
of the Boltzmann weight $\exp(-S)$, where $S$ is the action.
Gauge-invariant observables will remain intact provided
the following integral over any gauge orbit
\beq
  \int\cd\f\cd({\rm ghosts})\, \exp(-S(A_\m^\f,{\rm ghosts})) \,,
\label{orb}
\eeq
is a non-zero constant.
Here $\f(x)$ parameterizes a gauge transformation
and $A_\m^\f$ is the rotated field.

In perturbation theory, which is a saddle-point expansion around
the classical vacuum $A_\m=0$, this condition is satisfied.
Non-perturbatively, this is not necessarily true,
because of the existence of Gribov copies, namely, multiple
solutions of the gauge condition $\partial_\m A_\m^\f=0$
on the same orbit~\cite{GR}. In this situation the correct condition is
that $\sum {\rm sign}(\det(\O))$ must be a non-zero (integer) constant,
where the sum is over all Gribov copies on a given orbit.
Geometrically, one expects that that constant is the index of some
mapping~\cite{PH}.

In order to test this condition in a well-defined, non-perturbative
setting, one must resort to the lattice. Doing so, a no-go theorem was
discovered by Neuberger~\cite{HN}. Considering a
general class of BRST-invariant lattice theories, he proved,
under certain assumptions, that the orbit integral\seneq{orb}
is indeed a constant, but this constant is equal to zero!
Consequently the partition function itself, as well as unnormalized expectation
values of gauge-invariant operators, vanish.

Ways around Neuberger's theorem have been found.
The trick of \rcite{MT} is special to a U(1) gauge group,
and is therefore by itself of limited usefulness. Another approach,
where a non-abelian group is partially gauge-fixed to a maximal
abelian sub-group, was devised in \rcite{MS}. (It may be interesting to
combine these two methods into one gauge-fixing scheme.)

\section{Modified BRST transformations}

Here we will circumvent Neuberger's theorem by modifying
the ghost sector such that, when acting on the new ghost-sector
fields, BRST transformations cease to be nilpotent.
We are also guided by the consideration that
we would like to have
a {\it non-negative} ghost-sector partition function,
because positivity of the measure
is crucial for using existing numerical techniques.

Our starting point is the off-shell form
of the Faddeev-Popov gauge-fixing action
\beq
  S_{\rm FP} = {\tg^2\over 2} \l^2 + i\l\,\partial_\m A_\m + \bar{C}\O C \,,
\label{SFP}
\eeq
where $\l$ is the auxiliary field and $\tg^2=\x g^2$.
(We use continuum notation for simplicity, but
everything can be done on the lattice
in terms of the familiar compact link variables for the gauge field.)
We first perform a change of variables in the ghost sector.
Instead of the $\bar{C}$ field, we introduce a new grassmann field $\c$
and a real scalar field $\h$, both carrying a (suppressed) adjoint index.
At this point the gauge-fixing action is taken to be
\beq
  S_{\rm gf} = {\tg^2\over 2} \l^2 +  i\l\,\partial_\m A_\m + \c^T\O^T\O\, C
  + S_1 \,,
\label{Sgf}
\eeq
where
\bqry
  S_1
  &=&
  {h^2\over 2} \left( \brs(\O\c) +i\l \right)^2
\label{clown}
\\
  &=&
  \half \left( \O\h + h(\brs\O\c +i\l) \right)^2 \,.
\label{shift}
\eqry
Here $h$ is a new coupling constant.
$\brs$ is the BRST variation which, for the new field $\c$, is chosen
to be $\brs\c(x)=\h(x)/h$.

A comparison of  the old (\seeq{SFP}) and new (\seeq{Sgf}) gauge-fixing
actions reveals that, in effect, we have made the non-linear change
of variables $\bar{C} \to (\O\c)^T$. The necessary jacobian is
provided by the integral over the new $\h$ field. Indeed,
the gaussian $\h$-integration ``shifts away''
the $h$-dependent term in \seeq{shift}, resulting in a factor
of $|{\rm det}(\O)|^{-1}$. (Thus $h$ drops out of
the perturbative expansion.)
The grassmann-ghosts integral now yields
${\rm det}^2(\O)$. Putting these together we obtain (formally!)
that the ghost-sector partition function is $|{\rm det}(\O)|$.
Needless to say, perturbation theory is unchanged,
because in this context the sign of the determinant is inconsequential.

The familiar off-shell $\bar{C}$ transformation rule is $\brs\bar{C}=-i\l$.
For the new action, this relation is recovered as the
global minimum of the $\h$-action $S_1$, $\brs(\O\c)=-i\l$.
Moreover, the new action is BRST invariant
provided we choose $\brs\h(x) = - C(x)/h$. (For this to be true,
the $h$-dependent terms in \seeq{shift} are essential.)
Nilpotency is lost because $\brs^2\c(x) = - C(x)/h^2 \ne 0$.
In a non-abelian theory $\brs^2\h(x) \ne 0$ too, while $\brs^2$
still vanishes on all other fields.
Note that the mass dimension of all the ghost-sector fields is now zero.

\section{Gribov-horizon regulator}

In perturbation theory, the gauge-fixing actions\seneq{SFP} and\seneq{Sgf}
yield identical expressions for all correlation functions with no
external ghost legs. Non-perturbatively, however,
the $\h$-integral is ill-defined on the Gribov horizons,
where the Faddeev-Popov operator has zero modes.
To tame this singularity we add
a new {\it horizon-regulator} term to the action, $m^2 S_2$, where
\beq
  S_2 = \c^T D_\m^T D_\m C
  + \half \left( D_\m \h + h\, \brs({\rm Ad} A_\m)\c \right)^2
\label{hrz}
\eeq
and $(({\rm Ad} A_\m)\c)_a=f_{abc}A_{b\m}\c_c$.
The new addition\seneq{hrz} to the action is BRST invariant as well.
For certain (\eg Schr\"odinger functional) boundary conditions,
it can be shown that the lattice $\h$-action has no zero modes,
if $m\ne 0$. The lattice theory is therefore well-defined, and
BRST invariant. The target theory is defined by taking both
the continuum limit and the $m\to 0$ limit.

\section{Outlook}

We have constructed a new non-perturbative formulation of YM theory.
This gauge-fixed lattice theory is
not necessarily in the same universality class as the familiar lattice YM.
In perturbation theory, though, we can set $m=0$,
and so the two theories share an identical set of gauge-invariant correlation
functions to all orders. Proving unitarity
of the gauge-fixed theory beyond perturbation theory is an open question.

Non-perturbatively, the gauge-fixed partition function receives
new contributions for $m\to 0$.
They are $\d$-function-like distributions localized on the Gribov horizons
and proportional to $h$
(cf.\ eqs.~(\ref{shift},\ref{hrz});
we have verified this in a toy model~\cite{http}).
If we send $h \to 0$, these distributions vanish
and the ghost-sector partition function reduces to $|{\rm det}(\O)|$ exactly.
The double limit $m,h \to 0$ may thus provide a starting point
for numerical investigations using existing techniques.

The symmetries of the gauge-fixed theory include BRST
and a global remnant, $G$, of the local gauge group.
It will be interesting to explore the phase diagram and study
the realization of these symmetries in each phase.
An intriguing fact is that the dynamics of the longitudinal degrees of freedom
is controlled by the coupling constant $\tg$
which is {\it asymptotically free}, too (cf.\ \seeq{SFP});
this result follows from the known $\b$-functions of the gauge coupling
and the gauge parameter~\cite{free}. As a result, the gauge-fixed theory
has {\it two} dynamically-generated infra-red scales $\L$ and $\tilde\L$,
associated respectively with the renormalized coupling constants $g$ and $\tg$.

For $\L \gg \tilde\L$ we anticipate the existence of a phase with
unbroken symmetries, that may be identified with the familiar
confinement phase. We speculate that, for $\tilde\L \gg \L$, there may
exist a new phase where some of the global symmetries are
broken spontaneously.

Of particular interest would be a phase where the global
symmetry $G$ is broken spontaneously. Physically,
such a phase would represent a new, {\it dynamical} Higgs mechanism,
where (some of) the gauge bosons acquire a mass of order $\tilde\L$.
A necessary (and sufficient?) condition for the consistency of this phase
is that its low-energy description is given by a conventional,
renormalizable Higgs lagrangian. (Thus, our scenario might provide a new
solution to the Higgs-triviality problem.)
In the present context the effective Higgs field should be
a bound state made of glue, so it must be an adjoint Higgs.
Work on these and other issues is in progress.

{\it Acknowledgements}.
We thank the Inst.\ for Nuclear Theory at the Univ.\ of Washington
for hospitality.
This research is supported by the US -- Israel Binational
Science Foundation; MG is supported by the US Dept. of Energy.

\end{document}